\documentclass[prl,twocolumn,superscriptaddress]{revtex4}

\usepackage{amsmath,amsfonts,amssymb,amstext,amsthm,bbm}
\usepackage{graphicx}
\usepackage{times}

\def\>{\rangle}
\def\<{\langle}
\def\({\left(}
\def\){\right)}

\newcommand{\ket}[1]{|#1\>}
\newcommand{\bra}[1]{\<#1|}
\newcommand{\braket}[2]{\<#1|#2\>}

\DeclareMathOperator{\tr}{tr}

\newtheorem{theorem}{Theorem}

\begin{document}

\title{Most quantum states are too entangled to be useful as
computational resources}

\author{D.\ Gross}
\email{davidg@qipc.org}
\affiliation{ 
Institut f\"ur Mathematische Physik, Technische Universit\"at Braunschweig, 
38106 Braunschweig, Germany
}

\author{S.~T.\ Flammia}
\email{sflammia@perimeterinstitute.ca}
\affiliation{Perimeter Institute for Theoretical Physics, Waterloo, Ontario, N2L 2Y5 Canada} 

\author{J.\ Eisert}
\email{jense@qipc.org}
\affiliation{Physics Department, University of Potsdam, 14469 Potsdam, Germany}
\affiliation{Institute for Mathematical Sciences, Imperial College London, London SW7 2PE, UK}

\date{October ??, 2008}

\begin{abstract} 
It is often argued that entanglement is at the root of the speedup for
quantum compared to classical computation, and that one needs a
sufficient amount of entanglement for this speedup to be manifest. In
measurement-based quantum computing (MBQC), the need for a highly
entangled initial state is particularly obvious. Defying this
intuition, we show that quantum states can be too entangled to be
useful for the purpose of computation. We prove that this phenomenon
occurs for a dramatic majority of all states: the fraction of useful
$n$-qubit pure states is less than $\exp(-n^2)$, using concentration of
measure ideas.  Computational
universality is hence a rare property in quantum states. 
This work highlights a new aspect of the question concerning the role
entanglement plays for quantum computational speed-ups.  The
statements remain true if one allows for certain forms of
post-selection and also cover the notion of CQ-universality. We
identify scale-invariant states resulting from a MERA constructrion as
likely candidates for physically relevant states subject to this
effect.
\end{abstract}

\maketitle

\begin{figure}
  \centering
  \includegraphics[scale=.33]{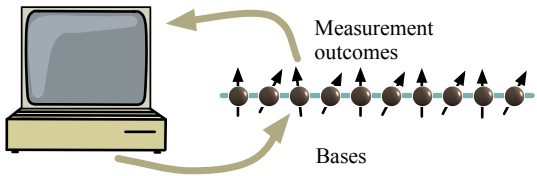}\\
  \caption{
    \label{fig:mbqcSchematics}
		The general setup of mea\-sure\-ment-based quantum computation.  A
		classical computer has access to a quantum state. In the course of
		the calculation, it may initiate arbitrary local measurements to
		be conducted on the state. The measurement outcomes are fed back
		into the computer, which can then use this data to decide
		upon future observables to be measured, and finally outputs the
		result of the computation
		\cite{Oneway,UniversalShort,UniversalLong,Janet}.
	}
\end{figure}

A classical computer endowed with the power to perform measurements on
certain entangled many-body states,  is strongly believed to be
exponentially more powerful than a classical machine alone (c.f.\
Fig.\ \ref{fig:mbqcSchematics}).  Indeed, a
computer 
having access to local measurements on a cluster state \cite{Oneway}
or the class of states identified in Refs.\ \cite{UniversalShort,
UniversalLong, Wires, Fundamentals,M} can efficiently simulate any
quantum computation.
The best-known classical algorithm for this task requires
super-polynomial run time and it is strongly believed that no
substantial improvement is possible. It is in this sense that certain
many-body states possess strong computational power. 
More precisely, the particular states mentioned above are
\emph{computaionally universal} in that they enable a classical
machine to effeciently solve any problem in the complexity class {\sc
BQP} \cite{FootnoteBQP}.

The key question that we ask in this work is: {\it How common is the
property of offering universal computational speed-ups and what is the
role of entanglement in this context?} 

All previous results which rule out computational universality of
certain quantum systems seem to do so by either (i) showing that the
systems are \emph{not entangled enough} to support a universal quantum
calculation \cite{MaartenResources, MaartenSimulation,VidalSimulation, ShiSimulation, FCS} 
or (ii) relying
on stringent symmetries \cite{GottesmanKnill, RichardGeneralizedStabs,
RichardMatchgates}.  It is therefore reasonable to conjecture, by
extrapolating from the current lines of research, that in generic
situations ``more entanglement'' will imply ``more computational
power''. 

Going further, it has been realized (using sundry techniques known
under the label of the 
``probabilistic method'' 
or the
``concentration of measure phenomenon'' 
\cite{Schechtman,Alon}), that generic quantum
states are extremely highly entangled from many points of view
\cite{HaydenRandomizing, HaydenGeneric, More}.  For example, a  typical state is almost
maximally entangled with respect to any partition of its systems into
two parties. It follows that most  states are excellent resources for
some quantum information protocols, e.g.\ teleportation with respect
to \emph{any} bipartition.  Thus, it is plausible to suspect that
offering a computational speed-up is a generic feature of quantum
states, if only advanced enough classical control schemes could be
devised to utilize their power.

{\it Main result. --} The arguments presented above turn out to be
fallacies. In a first step, we will show that families of states with
a large amount of entanglement -- as quantified by the \emph{geometric
measure} of entanglement \cite{ShimonyGeometric, LindenGeometric,
WeiGeometric} -- cannot be universal.  Recall that the geometric
measure  of a state vector $\ket\Psi$ is defined as 
\begin{equation*}
	E_g(\ket\Psi)=-\log_2 \sup_{\alpha \in \mathcal P}
	|\braket{\alpha}{\Psi}|^2, 
\end{equation*} 
where the supremum is taken over the set of all product states,
$\mathcal P$. In a second step, we proceed to demonstrate that our
criterion for large entanglement is fulfilled by typical quantum
states with overwhelming probability: they are too entangled to be
useful in this sense.  

The proof involves substituting the quantum resource by a fair coin.
In that sense, we show that even if one has complete knowledge about
the state used and is capable of designing the most sophisticated
measurement scheme, the distribution of the measurement outcomes is
not sufficiently different from that of a random string to
afford a universal speedup. This observation is the basis for the related results made independently in Ref.~\cite{BMW}.

The strategy of proof in the first step is as follows: We assume that
a classical computer assisted by local measurements on a highly
entangled state can efficiently identify both the solution to a
problem {\sc F} in {\sc NP} \cite{ComputationalComplexity}, and a
certificate for the solution. Under this assumption, we construct a
purely classical algorithm accomplishing the same task. Hence highly
entangled states cannot cause a significant speed-up for these
particular problems. For concreteness, one may think of {\sc F} as the
paradigmatic {\sc Factoring} problem: given an integer $N$ and an
interval $[k,l]$, decide whether $N$ has a factor contained in
$[k,l]$. A certificate for the solution is provided by the prime
decomposition of $N$. Since there is an efficient quantum
algorithm identifying the prime decomposition \cite{shor}, it follows
that highly entangled states cannot be universal (unless {\sc
Factoring} is in {\sc P}, which is generally believed to be highly
unlikely).

\begin{theorem}[Uselessness of quantum states for MBQC]\label{Thm:useless}
	Let $\ket{\Psi_n}$ be an $n$ qubit state with geometric measure of
	entanglement $E_g(\ket{\Psi_n})>n-\delta$. Consider a classical computer 
	augmented by the power to perform local measurements on $\ket{\Psi_n}$.
	Assume this joint system is capable 	of finding and certifying a
	solution to an NP problem {\sc F} after $t$ time steps, with
	probability of success at least $1/2$.  Then there exists a
	\emph{purely classical} algorithm which identifies a solution to
	{\sc F} after $C(n) 2^{\delta+1}\ln(1/p_f)$ time steps with
	probability of success at least $1-p_f$.  Here, $C(n)$ is the time
	it takes to verify the certificate on a classical computer.
\end{theorem}

Note that $C$ is a polynomial function of $n$ (this being the defining property of NP problems). The theorem implies that a family of states $\ket{\Psi_n}$ cannot provide a super-polynomial speedup whenever their geometric measure is of the form $E_g(\ket{\Psi_n})=n-O(\log_2 (n))$. A priori it is unclear that states with such an extreme geometric entanglement exist at all.  It turns out that not only do they exist, but that this property is shared by the vast majority of all many-body states.

\begin{theorem}[Almost all states are useless]\label{thm:almost}
	The fraction of state vectors on $n\geq11$ qubits with geometric measure of 
	entanglement less than $n-2\log_2 (n)-3$ is smaller than $e^{-n^2}$.
\end{theorem}

It immediately follows that the fraction of universal
resources among $n$ qubit pure states is less than $e^{-n^2}$.

To prove Theorem 1, we assume that the classical part of the algorithm is deterministic, which does not restrict generality, since any probabilistic parts may be implemented by using quantum measurements as coins.  In the course of the calculation, the computer will perform up to $n$ local single-qubit projective measurements with two outcomes each, obtaining one of $2^n$ possible sequences of outcomes.  There is a set $G$ of ``good'' outcomes, which will cause the computer to output a valid solution to the problem {\sc F} after $t\leq n$ time steps. By assumption, the probability of obtaining an outcome from $G$ is larger than $1/2$. Each element of $G$ is labeled by a product state $\ket\alpha$ corresponding to the local measurement outcomes. The probability of the event associated with $\ket\alpha$ to occur is 
$|\braket{\alpha}{\Psi}|^2 \leq 2^{-E_g\left(\ket\Psi\right)}
\leq 2^{-n+\delta}$.
Hence
\begin{eqnarray*}
	1/2 \leq \operatorname{Prob}(G) < |G|\,
	2^{-n+\delta}
	&\Rightarrow& 
	|G| > 
	2^{n-\delta-1}.
\end{eqnarray*}
Thus the ratio of good outcomes to the total number obeys
\begin{equation}\label{ratio}
	|G|/2^n > 2^{-\delta-1}.
\end{equation}	

To simulate the procedure on a classical computer, use the following algorithm: Instead of performing a physical measurement, choose the outcome of the measurements randomly using a fair coin, to generate a random string. This string is fed into the same classical postprocessing algorithm as before. If the random string causes the classical part of the computation to output a result after $t$ time steps, check whether it solves the problem {\sc F}.  The problem being in NP, this is efficiently possible. If the result is valid, output it and abort. Otherwise 
repeat the procedure with another random string. The probability of not having obtained a valid outcome after $k$ trials is bounded above by 
$(1-2^{-\delta-1})^k < e^{-k\,2^{-\delta-1}}$, using Eq.\ (\ref{ratio}) of the cardinality of the set of ``good'' outcomes. Set $k= 2^{\delta+1}\ln(1/p_f) $ to achieve a probability of failure smaller than $p_f$. The claim of Theorem 1 is now immediate. $\Box$

The proof of Theorem 2 requires two technical ingredients. The first is a concentration of measure result: Let $\ket\Phi$ be a normalized vector in $\mathbb{C}^d$, and let $\ket\Psi$ be drawn from the unit sphere according to Haar measure. Then
\begin{equation}\label{eqn:concentration}
	\operatorname{Prob}\{
		|\braket{\Phi}{\Psi}|^2\geq\varepsilon
	\}
	< \exp(- (2d-1) \varepsilon).
\end{equation}
This statement follows easily from standard bounds to be found e.g.\
in Refs.\ \cite{HiaiPetz, petzHaar}. Secondly, we require the concept
of an $\varepsilon$-net \cite{Schechtman, HaydenGeneric, HaydenRandomizing}. An
\emph{$\varepsilon$-net $\mathcal{N}_{\varepsilon,k}$ on the set
$\mathcal P$ of product states on $k$ qubits} is a set of vectors such
that
\begin{equation}\label{eqn:net}
		\sup_{\alpha \in \mathcal P}\inf_{\tilde\alpha\in\mathcal{N}_{\varepsilon,k}}
		\big\|\ket\alpha-\ket{\tilde\alpha}\big\|<\varepsilon/2.
\end{equation}
We claim such a net exists whose cardinality is bounded by
	$|\mathcal{N}_{\varepsilon,k}|\leq(5k/\varepsilon)^{4k}$.
Indeed, from Ref.\ \cite{HaydenRandomizing} we know that there is an $(\varepsilon/k)$-net $\mathcal{M}$ on the space of single qubit state vectors, where $|\mathcal{M}|\leq (5k/\varepsilon)^4$. Set 
\begin{equation*}
	\mathcal{N}_{\varepsilon,k}=\{
	\ket{\tilde \alpha_1}\otimes\dots\otimes\ket{\tilde \alpha_k} \,:\,
	\ket{\tilde \alpha_i}\in\mathcal{M}\}.
\end{equation*}
Now let $\ket{\alpha}=\bigotimes_i^k \ket{\alpha_i}$ be a product vector.  By definition of $\mathcal{M}$, for every $i$ there exists $\ket{\tilde\alpha_i}\in\mathcal{M}$, such that $|\braket{\alpha_i}{\tilde\alpha_i}|^2\geq 1-\varepsilon^2/{4k}$. Hence, for $\ket{\tilde \alpha}=\bigotimes_i\ket{\tilde\alpha_i}$, 
\begin{equation*}
	|\braket{\alpha}{\tilde\alpha}|^2
	\geq
	\left(1-\frac{\varepsilon^2}{4k}\right)^k
	\geq
	1-\frac{\varepsilon^2}{4},
\end{equation*}
which implies Eq.~(\ref{eqn:net}) \cite{HaydenRandomizing}.

Proceeding to prove Theorem 2, we let $\varepsilon=2^{-l}$ for some yet-to-be-determined number $l$. Let $\mathcal{N}_{\varepsilon,n}$ be as above. Employing the standard union bound, we find
\begin{eqnarray}
	&&\operatorname{Prob}\{
		\sup_{\ket{\tilde\alpha}\in\mathcal{N}_{\varepsilon,n}}
		|\braket{\tilde\alpha}{\Psi}|^2\geq 2^{-l}
	\} \nonumber\\
	&<& 
	\exp(- (2^{n+1}-1) 2^{-l} )
	\,\big|\mathcal{N}_{\varepsilon,n}\big| \nonumber\\
	&<& 
	\exp(- 2^{n-l} + 2nl \ln (2) + 4n\ln(5n) ) \nonumber\\
	&<& 
	\exp(- 2^{n-l} + 2nl ) \label{eqn:fromSomeN}\\
	&<&
	\exp(-2^{n-l}+2n^2) \label{eqn:intermediateProb}
\end{eqnarray}
where the estimate (\ref{eqn:fromSomeN}) is valid if
\begin{equation}\label{eqn:lCondition}	
	2nl(1-\ln(2))>4\ln(5n).
\end{equation}
Choosing $l=n-\log_2(3n^2)$, the condition above is satisfied when $n\geq11$. Further, Eq.\ (\ref{eqn:intermediateProb}) becomes $\exp(-n^2)$. Now let $\ket\alpha$ be a general product vector and $\ket{\tilde\alpha}$ be the closest element in the $\varepsilon$-net. Then, using the operator norm $\|.\|_\infty$,
\begin{eqnarray*}
	&&
	\big|
	|\braket{\alpha}{\Psi}|^2-
	|\braket{\tilde\alpha}{\Psi}|^2
	\big|
	=
	\big|
	\tr\big[
		(\ket{\alpha}\bra{\alpha}-\ket{\tilde\alpha}\bra{\tilde\alpha})
		\ket\Psi\bra\Psi
	\big]
	\big|\\
	&\leq&
	\big\|	
		\ket{\alpha}\bra{\alpha}-\ket{\tilde\alpha}\bra{\tilde\alpha}
	\big\|_\infty
	\leq
	\big\|	
		\ket{\alpha}\bra{\alpha}-\ket{\tilde\alpha}\bra{\tilde\alpha}
	\big\|_1
	\leq\varepsilon=2^{-l}.
\end{eqnarray*}
Here we have used that for the trace-norm $\|.\|_1$,
\begin{equation*}
	\big\|	
		\ket{\alpha}\bra{\alpha}-\ket{\tilde\alpha}\bra{\tilde\alpha}
	\big\|_1\leq 2 
	\big\|\ket\alpha-\ket{\tilde\alpha}\big\|
\end{equation*}
(see, e.g., Ref.~\cite{HaydenRandomizing}). It follows that
\begin{eqnarray*}	
	\sup_{\alpha\in \mathcal{P}} |\braket{\alpha}{\Psi}|^2
	\leq 2^{-l+1}
	<2^{-n+2\log_2 (n)+3}
\end{eqnarray*}
with probability greater than $1-e^{-n^2}$. $\Box$

{\it CQ-universality and PostBQP. --} 
References \cite{Fundamentals,MaartenResources} introduced a more
stringent benchmark for universal resource states.
The authors ascribe the quality of ``CQ-universality'' to a resource
state if---up to local unitary corrections---any pure state on $k$
qubits can be prepared out of a sufficiently large $n>k$ qubit
resource by means of local measurements on the remaining $n-k$ sites.
As has been shown in Ref.\ \cite{UniversalShort,UniversalLong}, efficient
CQ-universality is a strictly stronger requirement than the notion of
universality used in
the present paper.  Hence, the results presented above already imply
that generic states are not CQ-universal. However, we can strengthen
the statement: Most states fail to be CQ-universal, even if we assume
we had the power to \emph{choose\/} the local measurement outcomes. 

To put this into perspective, recall that models of quantum computing
with the assumed capability of choosing the outcome of at least one
measurement have been analyzed under the label of \emph{post-selected}
quantum computing \cite{PostBQP}. The complexity class of decision
problems efficiently decidable on post-selected computers is {\sc
PostBQP}. It is known \cite{PostBQP} that {\sc PostBQP} $=$ {\sc
PP} $\supseteq$ {\sc NP} -- which implies that the capability to
postselect dramatically increases the computational strength of
quantum computers (unless {\sc PP $\subseteq$ BQP}). 

For generic states in the CQ-setting, however, postselection does not
seem to help. Indeed, the dramatic majority of states are not efficiently
CQ-universal, even if we 
(i) assume the power to choose the outcome of the local measurements, 
(ii) allow for any fixed error $\varepsilon$ in the output fidelity,
(iii) ask only for the capability to prepare a single product state,
and
(iv) make no a priori assumption on which sites the final state should
end up in.

We proceed to prove the claim. Fix a subset $K$ of $k$ sites (the
``output register'') of a given n-qubit resource state $\ket\Psi$.
Combining Lemma III.5 in Ref.\ \cite{HaydenGeneric} with Theorem 2 above, we
find that for a random resource $\ket\Psi$ and fixed $\varepsilon > 0$,
\begin{eqnarray*}
	\sup_{\alpha,\beta}
	\ln
	\frac{|(\bra{\alpha}\otimes\bra{\beta})\ket{\Psi}|^2}
	{\|\braket{\beta}{\Psi}\|^2} \nonumber 
	<
	2\ln n -k - \ln(1-\varepsilon) + 3 \label{eqn:cqFid}
\end{eqnarray*}
with probability at least $1-e^{-n^2}+e^{-c 2^k \varepsilon^2}$.
The supremum is over product states $\ket\alpha$ on $K$ and $\ket\beta$ on
$K^C$.
There are fewer than $n^k$ choices for
$K\subset\{1,\dots,n\}$ of cardinality $k$. Hence the preceeding bound
is true for all such $K$ simultaneously with probability no smaller
than
\begin{equation*}
	1-\big(e^{-n^2+k\ln n}+e^{-c 2^{k} \varepsilon^2+k\ln n}\big).
\end{equation*}
The claim follows by taking $n\to\infty$ and $n=\operatorname{poly}(k)$.

{\it Efficiently preparable states and concrete examples. --} While
conceptually relevant, generic Haar-random states on large systems
cannot be efficiently prepared. In this section, we demonstrate that
the effect that ``too much entanglement'' impedes computational
efficiency can also be identified in more physically relevant states.

First, we exhibit a family of efficiently preparable states
$\ket{\Phi_n}$ on $n$ qubits, for which $\lim_{n\to\infty} E_g(|\Phi_n\rangle)
/ n =1$. These values of the geometric measure are high enough to rule
out the possibility that such families offer an exponential speedup
for the kind of problems considered above.
Constructing explicit states which cannot cause even a
super-polynomial increase in computational power remains an open
problem \cite{improvements}.

\begin{figure}
  \centering
  \includegraphics[scale=.33]{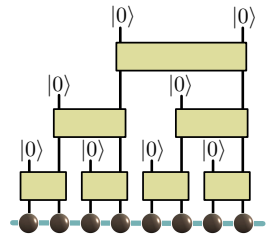}\\
  \caption{
    \label{fig:mera}
		The tree-tensor network or instance of a MERA
		\cite{MERA,ShiSimulation,MaartenSimulation} that gives rise to
		states with high geometric entanglement. Each of the boxes
		represents a unitary.}
\end{figure}
	
The states are related to the MERA construction introduced in
Ref.~\cite{MERA} in order to capture the properties of critical, scale
invariant states. Fix a dimension $d$ and choose a unitary $U\in
U(\mathbbm{C}^d)$. Define $V:
\mathbbm{C}^d\to\mathbbm{C}^d\otimes\mathbbm{C}^d$ by $V\ket\beta =
U\ket0\otimes\ket \beta$. For every $k$, we construct the state
$\ket{\Phi_k}$ on $n=2^k$ qudits by concatenating the maps $V$ in a
tree-like fashion as shown in Fig.~\ref{fig:mera}. Now set
$d=\operatorname{poly}(k)$ and choose $U_k\in U(\mathbbm{C}^d)$
randomly for
every $k$. Recall that, by the Solovay-Kitaev theorem \cite{SK}, general unitaries
in $U(\mathbbm{C}^m)$ can be efficiently approximated by a quantum
circuit in a time polynomial in $m$ and polylogarithmic in the error.
Employing once more
standard concentration of measure arguments \cite{HaydenGeneric}, one easily
finds that
\begin{equation*}
	\lim_{k\to\infty}
	\sup_{\alpha_1,\alpha_2,\beta} |\bra{\alpha_1,\alpha_2} V_k
  \ket{\beta}|^2 = 1/d(k).
\end{equation*}
Thus, $V_k^\dagger$ maps any product state
$\ket{\alpha_1}\otimes\ket{\alpha_2}$ to a vector $\ket{\beta}$ 
with $\|\ket\beta\|^2\leq
1/d(k)$ (asymptotically). It follows that the lowest layer of
$V^\dagger$'s in
the definition of
$\ket{\Phi_k}$ sends any unit-norm product vector on $n$ qudits to a
product vector on $n/2$ qudits, with squared norm at most 
$d(k)^{-n/2}$. Inducting over all layers and using the appropriate
base-$d$ logarithm in the definition of geometric measure for
qudits, we conclude \cite{improvements}
\begin{eqnarray*}
	\lim_{k\to\infty} \frac{-1}{n(k)} \log_{d(k)} 
	\sup_{\alpha\in\mathcal{P}}|\braket{\alpha}{\Phi_k}|^2
	=\sum_{i=1}^\infty 2^{-i}=1 \ .
\end{eqnarray*}

Turning to the CQ-setting, it is easy to identify states whose
efficiency as a resource is limited by the presence of correlations
which can, on average, not be removed by means of local measurements.
Consider the ground state $|\Psi\rangle$ of a
{\it critical model} with two-point correlation functions between the sites
$j,k$ fulfilling 
\begin{equation*}
	\langle \Psi | M_j \otimes M_k | \Psi\rangle - 
	\langle \Psi | M_j | \Psi\rangle
	\langle \Psi |  M_k | \Psi\rangle
	> f(\text{dist}(j,k)),
\end{equation*}
with $f(x)=1/\operatorname{poly}(x)$. Assume we could prepare a
three-qubit cluster state $|Cl_3\rangle\langle Cl_3|$ on some ``output
register''. We allow for local
unitary corrections, and an average trace-norm error of $\varepsilon$,
so that, with
\begin{eqnarray*}
	\sigma=
	p_l U_j^{(l)} \otimes  U_i^{(l)} \otimes U_k^{(l)}  |
	Cl_3 \rangle\langle Cl_3 | (U_j^{(l)} \otimes  U_i^{(l)} \otimes
	U_k^	{(l)})^\dagger\nonumber
\end{eqnarray*}
it holds that 
\begin{equation}
	\|\text{tr}_{\backslash\{j,i,k\}}(
	|\Psi\rangle\langle\Psi|)-\sigma\|_1\leq \varepsilon.
\end{equation}	
Here, $p_l$ is the probability of obtaining a certain sequence of
measurement outcomes. Noting that the qubits associated with sites
$j,k$ are uncorrelated for $\ket{Cl_3}$, one easily derives that the
size of the resource must scale as $O(f^{-1}(\varepsilon))$. This
observation complements the results of Ref.\ \cite{Fundamentals},
where different arguments for the fact that critical ground states may
not be well-suited as CQ-resourcess were presented.

{\it Summary and outlook. --} We have shown that entanglement for
universal resource states must ``come in the right dose''. Future work
should aim to identify a greater variety of physically relevant states
exhibiting the phenomenon of being ``too entangled''. For example, it
would be intersting to quantify to which degree output states of
random, polynomially sized quantum circuits are subject to this
effect. Also, the results underscore the importance of systematically
understanding relevant classes of the few states that are in fact
universal. This work highlights the quite intriguing role
entanglement plays in quantum computing. As with most good things, it
is best consumed in moderation. 

\emph{Acknowledgments --} The authors thank M.\ Bremner, C.\ Mora, and
A.\ Winter for discussions. After this work had been completed, we
learned about related results in preparation \cite{BMW}. DG and JE
thank the EU (COMPAS, CORNER, QAP), the EPSRC, the EURYI, and
Microsoft Research for support.  STF was supported by the Perimeter
Institute for Theoretical Physics.  Research at Perimeter is supported
by the Government of Canada through Industry Canada and by the
Province of Ontario through the Ministry of Research~\& Innovation.

\section{Appendix -- Geometric measure in graph states}

{\it Graph states} \cite{GS} constitute the prototypical resource for
one-way computation \cite{Oneway, Fundamentals}. We will briefly
sketch methods for bounding the geometric measure of entanglement for
graph states, and relate it to a measure with more operational
meaning: the {\it Pauli persistency} \cite{GS}. The latter quantity is
given by the minimal number of local Pauli measurements that are needed
to deterministically completely disentangle a quantum state. 

Consider a distinguished system, which we label $1$ without loss of generality.
In terms of the eigenvectors of a local Pauli
operator at site $1$, we can write
\begin{equation*}
	|\Psi\rangle = \left(| y_1\rangle |\psi\rangle 
	+| y_1'\rangle 
	|\psi'\rangle \right)
	2^{-1/2}.
\end{equation*}
Let $|\tilde \alpha\rangle=|\alpha_2\rangle\otimes \dots \otimes
|\alpha_n\rangle$ be a product vector maximizing the overlap
with $\ket\psi$.
It is known that $| \psi\rangle$ and
$| \psi\rangle'$ are local unitarily equivalent \cite{GS}. Therefore
$E_g(|\psi\rangle) = -\log_2|\langle \tilde \alpha |
	\psi\rangle|^2 = E_g(|\psi'\rangle)$.
We hence have that
\begin{eqnarray*}
	E_g(|\Psi\rangle) \leq 
	-\log_2|\langle y_1|\langle \tilde \alpha | 
	\Psi\rangle|^2\nonumber
	=
	E_g(|\psi\rangle) + 1,
\end{eqnarray*}
implying 
$E_g(|\Psi\rangle) \leq 
	\frac{1}{2}E_g(|\psi\rangle)+ 
	\frac{1}{2}E_g(|\psi'\rangle)
 + 1$.
Iterating this argument, we find that if $k$ local Pauli measurements
are sufficient to completely disentangle a graph state vector
$|\Psi\rangle$ 
then $E_g(|\Psi\rangle )\leq k$. 
(Interestingly, an identical bound also holds for the {\it Schmidt
measure} \cite{GS}.)
Conversely, a lower bound to the geometric measure is given by $E_g(
|\Psi\rangle ) \geq\log_2 \text{rank}_{\mathbbm{F}_2} (\Gamma')$,
where $\Gamma'$ is the off-diagonal sub-matrix of the adjancency
matrix $\Gamma$ of the graph state, and the rank is taken over
$\mathbbm{F}_2$. 

Note that both estimates may be formulated solely in terms of the
efficient description of $\ket\Psi$ by its graph and can be
calculated explicitely for many such states. We are presently not
aware of a family of graph states which is ``too entangled'' in the
sense introduced above.


\begin{thebibliography}{99}
\bibitem{FootnoteBQP}		
	{\sc BQP} is the set of decision problems solvable by a
	polynomial-sized uniform quantum circuit with bounded error probability.
\bibitem{improvements}
	There are several obvious ways for improving the bounds.
	Theorem 1 uses the geometric measure only to bound the 
	``number of product states needed to account for half the norm of
	the resource''---a quantity which could be bounded directly. Also,
	one could choose the $V_k$'s such that their range lies in the
	antisymmetric subspace. Then, if one uses the optimal product
	states for the l.h.s.\ of any branch, one would be forced to choose
	different, non-optimal states on the r.h.s..
\end{thebibliography}
\end{document}